\documentclass[english,twocolumn]{revtex4-1}

\pdfoutput=1

\usepackage[T1]{fontenc}
\usepackage[latin9]{inputenc}
\usepackage{geometry}
\usepackage{booktabs}
\usepackage{bm}
\usepackage{amsmath}
\usepackage{amssymb}
\usepackage{graphicx}
\usepackage{wasysym}
\usepackage{esint}

\makeatletter

\providecommand{\tabularnewline}{\\}
\newcommand{\lyxdot}{.}

\newcommand{\angstrom}{\mbox{\normalfont\AA}}

\makeatother

\usepackage{babel}
\begin{document}

\title{Fluctuation capture in non-polar gases and liquids}

\author{D. G. Cocks, R. D. White}

\affiliation{College of Science, Technology and Engineering, James Cook University,
Townsville 4810, Australia}
\begin{abstract}
We present a new model to identify natural fluctuations in fluids,
allowing us to describe localization phenomena in the transport of
electrons, positrons and positronium through non-polar fluids. The
theory contains no free parameters and allows for the calculation
of capture cross sections $\sigma_{\mathrm{cap}}(\epsilon)$ of light-particles
in any non-polar fluid, required for non-equilibrium transport simulations.
We postulate that localization occurs through large shallow traps
before stable bound states are formed. Our results allow us to explain
most of the experimental observations of changes in mobility and annihilation
rates in the noble gases and liquids as well as make predictions for
future experiments. Quantities which are currently inaccessible to
experiment, such as positron mobilities, can be obtained from our
theory. Unlike other theoretical approaches to localization, the outputs
of our theory can be applied in non-equilibrium transport simulations
and an extension to the determination of waiting time distributions
for localized states is straight forward.
\end{abstract}
\maketitle
\emph{Introduction---}The transport of light particles (i.e. electrons
\textbf{$e^{-}$}, positrons $e^{+}$ or positronium Ps) through materials
takes place in a wide range of systems, occurring in plasma research,
medical imaging, particle therapy, organic solar cells and particle
detectors. Until recently, it was typical for transport in dense fluids
(i.e. liquids and dense gases) to be modeled using the same assumptions
as dilute gases, including effects of the medium only through terms
linear in the bulk density, neglecting correlations between the fluid
particles and assuming independent collisions between the light and
fluid particles. Improvements to traditional transport theory have
made use of the seminal works by Cohen and Lekner~\citep{Cohen1967}
to partially address the structure of a dense fluid~\citep{Boyle2015}
which explain order of magnitude changes in observed drift velocities
and negative-differential-conductivities with an applied electric
field.

Correlations in a dense structured fluid also introduce new processes
such as ``bubble capture''~\citep{Iakubov1982,Hernandez1991},
whereby a light-particle (LP) strongly interacts with the surrounding
fluid to form a stable bound state, analogous to the polaron of solid-state
physics. These bound states consist of a reduction or enhancement
of the local density surrounding the light-particle, leading to ``bubbles''
or ``clusters'', respectively. The combined fluid/light-particle
object can possess an effective mass much greater than the quasi-free
particles leading to modified transport coefficients and even fractional
diffusion~\citep{Philippa2014}. For $e^{+}$ and Ps, the change
in density surrounding the particle leads to significant enhancements
or reductions in the annihilation rate and hence the observed lifetime
of the particles.

This paper presents a new model for fluctuation capture which can
be applied to all non-polar fluids without resorting to empirical
inputs. To the knowledge of the authors, this model is the first to
produce energy-dependent capture cross sections $\sigma_{\mathrm{cap}}(\epsilon)$,
along with details sufficient to calculate energy-dependent waiting-time
distributions $\theta(\epsilon,t)$. This enables the inclusion of
fluctuation capture into non-equilibrium ab~initio transport simulations.

There have been many indirect experimental observations of fluctuation
capture and we choose to focus here on experiments in the noble gases.
For $e^{-}$, there have been observations of reduced mobility in
dense helium and neon gases~\citep{Bartels1975,Jahnke1973,Borghesani1992},
as well as fractional diffusion in liquid neon~\citep{Sakai1992},
and suggestions of fluctuation capture in xenon~\citep{Huang1978}.
For $e^{+}$ and Ps, significant changes in annihilation rates have
been observed for helium, neon, argon and xenon~\citep{Canter1975,Canter1975a,Rytsola1984,Pepe1995,Nieminen1980,Hernandez1991,Tuomisaari1999}
beyond the linear dependence on density expected in the dilute gas
phase. Ps bubbles, which dramatically increase the Ps lifetime, have
also been observed through the use of ACAR measurements in argon,
krypton and xenon~\citep{Varlashkin1971,Coleman1999}.

Many theoretical investigations into fluctuations have focused only
on the equilibrium state of the dressed particle~\citep{Iakubov1982,Hernandez1991,Miller2002}.
In this paper, we are interested in an inherently non-equilibrium
and dynamical process. Some previous investigations for light particles
out of equilibrium have looked at ``cavities'' or ``voids'' within
fluids, considering the largest available volume with a total absence
of atoms \citep{Sakai1992,Mikhin2005,Schnitker1986,Vishnyakov2000,Shen1999,Stepanov2012}.
Far less prevalent are investigations into mesoscopic fluctuations
with a weak perturbation of the density profile \citep{Eggarter1971,Space1992,Truskett1998,English2011}.
We emphasize that our focus here is on fluctuation capture of a light
particle and so do not address the formation of Ps, which is a complex
process~\citep{Stepanov2012}. 

Despite these previous investigations, we believe that a significant
omission remains in the literature for fluctuation capture. As the
de~Broglie wavelength of a thermalized LP can extend to over hundreds
of average fluid particle separations, it is disconnected with the
picture of a tight compact void which is of the order of a single
spacing. Many of the theoretical studies referenced above either assume
the light-particle is already in the void, or invoke a classical description
of the particle, at odds with the quantum nature dictated by the energy
and length scales of the system. Instead we believe that a larger
mesoscopic scale fluctuation is necessary to describe the leading
contributions to localization and the stable trapped state results
from a contraction of these large-scale fluctuations rather than expansion
of small-scale voids. Furthermore, we propose a general theory that
can be applied to any material in the gas or liquid phases without
requiring free parameters. We have applied this theory to the noble
gases and liquids, successfully explaining almost all of the current
experimental measurements and predict transport and localization behavior
for combinations of atomic species and densities which have not yet
been measured.

The energy-dependent capture cross-sections $\sigma_{\mathrm{cap}}(\epsilon)$
and waiting time distributions $\theta(\epsilon,\tau)$ obtained from
this model can be used directly in solutions of the Boltzmann equation~\citep{Philippa2014}
and Monte-Carlo transport simulations.  The inputs to our calculations
are the fluid interparticle potential, the elastic cross sections
for the light-particle/fluid interaction and certain hydrodynamic
properties of the fluid. We choose to model the noble gases in their
gaseous and liquid states by an untruncated Lennard-Jones (LJ) fluid
with the appropriate values for the energy ($\epsilon_{\mathrm{LJ}}$)
and length ($\sigma_{\mathrm{LJ}}$) scales detailed in the supplementary
material. The LJ parameters also provide a useful unit, $T_{\mathrm{LJ}}=\epsilon_{\mathrm{LJ}}/k_{B}$,
for temperature. The calculation proceeds in four steps: identifying
the fluctuations, calculating their properties, obtaining binding
energies and calculating a rate of capture.

\begin{figure}
\begin{centering}
\includegraphics[width=0.24\textwidth]{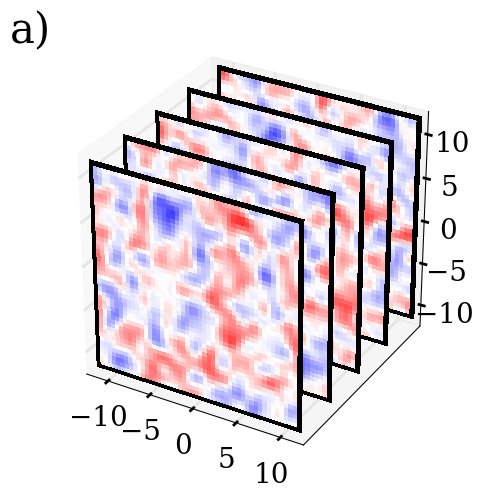}\includegraphics[width=0.24\textwidth]{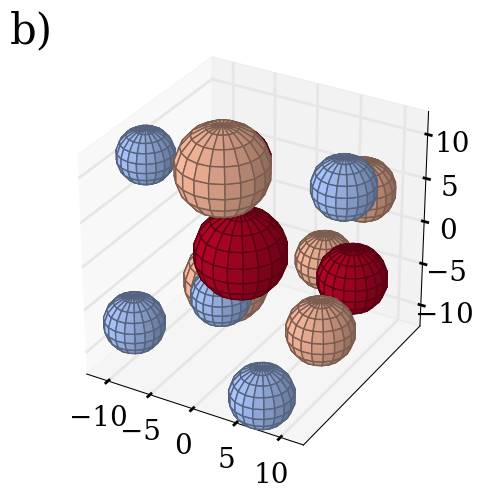}
\par\end{centering}

\caption{\label{fig:identification-process}Steps in the fluctuation identification
process for a single snapshot of a system with $T=1.0\,T_{\mathrm{LJ}}$,
$\rho=0.71\,\rho_{\mathrm{LJ}}$. a) Cross-sections of the continuous
density distribution obtained by Gaussian blurring the atomic positions.
b) The identified fluctuations, represented as spheres whose radius
is $R_{c}$ and whose threshold $t_{c}$ is given by the color, with
blue (red) as the most deep (shallow) fluctuations.}
\end{figure}

\emph{Fluctuation identification---}It is first necessary to understand
the distribution of fluctuations and their spatial profiles in the
absence of the light particle, which we inevstigate by performing
Monte-Carlo simulations of the LJ fluid, saving snapshots from the
simulation. From these snapshots we create a continuous density distribution
by blurring the atomic positions with Gaussians of width $\sigma_{\mathrm{blur}}=(3/4\pi\rho_{0})^{1/3}$,
corresponding to the Wigner-Seitz radius. An example of this blurred
distribution is shown in figure~\ref{fig:identification-process}a).
We then take averages over the continuum density in spherical volumes
throughout the fluid, gradually increasing the radii, $R$, of the
volumes. A fluctuation will have an average density that deviates
significantly from the bulk value even after averaging over a large
region. For each averaging volume we look for connected regions with
average densities that fall above or below a set of prescribed thresholds
$t_{c}$. A high-(low-)density fluctuation is identified as the largest
radius, $R_{c}$, that is above (below) the threshold for which a
connected region can still be found. A sample of identified fluctuations
is shown in figure~\ref{fig:identification-process}b).

\begin{figure}
\centering{}\includegraphics[width=0.48\textwidth]{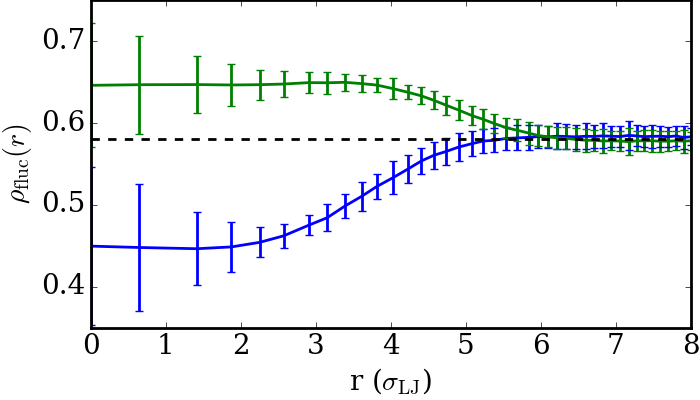}\caption{\label{fig:Density-distributions}Mean density distributions, $\rho_{\mathrm{fluc}}(r)$
for a system with $T=1.2T_{\mathrm{LJ}}$, $\rho=0.58\rho_{\mathrm{LJ}}$
and a low-(high-)density fluctuation classified as $t_{c}=0.84$,
$R_{c}=3.97$ ($t_{c}=1.1$, $R_{c}=4.99$) with one standard deviation
in the statistical average shown. The large deviations for small $r$
are expected, as these are sensitive to the exact positions of the
atoms.}
\end{figure}

\emph{Fluctuation properties---}For each fluctuation classified by
the parameters $\{R_{c},t_{c}\}$ we determine the density distribution
spherically averaged about its center point, $\rho_{\mathrm{fluc}}^{\{R_{c},t_{c}\}}(r)$
where $r$ is the radial distance from the center of the fluctuation.
To build up a good representation of a macroscopic liquid, we perform
a statistical average over $\rho_{\mathrm{fluc}}^{(R_{c},t_{c})}(r)$
for many different uncorrelated snapshots of the Monte-Carlo simulation.
 We show some example density distributions, $\rho_{\mathrm{fluc}}^{(R_{c},t_{c})}(r)$,
in figure~\ref{fig:Density-distributions}. During the statistical
averaging, we also extract a density of states for the fluctuations
of the same threshold, $g_{t_{c}}(R_{c})\,dR_{c}$, by binning the
counts of the fluctuations into a coarse-grained grid of $R_{c}$.

The two quantities, $\rho_{\mathrm{fluc}}^{\{R_{c},t_{c}\}}(r)$ and
$g_{t_{c}}(R_{c})$ are what sets this theory apart from others. These
quantities are essential to allow the connection to macroscopic non-equilibrium
transport simulations which require energy-resolved collision frequencies.

\emph{Binding energies and capture rate---}We use the density profiles
$\rho_{\mathrm{fluc}}^{(R_{c},t_{c})}(r)$ to calculate the probability
that the fluctuation will capture an incoming light particle by performing
a scattering calculation, whereby the initial quasi-free LP transitions
to a bound state within the fluctuation, which requires us to define
a scattering potential $V(r)$ and a coupling $W(\epsilon\rightarrow\epsilon_{b})$.
The effective interaction between the LP and the bulk, $V(r)$, is
chosen in a homogeneous-energy local-density (HELD) approximation.
The term ``homogeneous energy'' refers to the background energy
felt by a quasi-free particle in a homogeneous system, $V_{0}(\rho_{0})$,
which has a complicated non-linear dependence on bulk density~\citep{Nazin2008}.
We connect this to the effective interaction by assuming that the
LP feels the local density only, i.e. $V(r)=V_{0}(\rho(r))$. This
``local approximation'' neglects effects from the spatial-dependence
of screening (note, however, that non-local effest of microscopic
screening are included within $V_{0}$) which is negligible for the
small density changes we consider, and has fared very well in previous
calculations~\citep{Nazin2008}. The sign of $V_{0}$ determines
the types of fluctuations that are favored by the light particle:
low-density bubbles or high-density clusters for $V_{0}>0$ and $V_{0}<0$
respectively.

In this paper, we choose a simple representation of $V_{0}$ for the
sake of clarity. Instead of the full non-linear dependence, we take
$V_{0}(\rho_{0})\approx2\pi\hbar^{2}a_{s}\rho_{0}/m$ where the scattering
length $a_{s}$ describes a Born-like approximation for the LP-atom
interaction. We emphasize that we make this choice only so that we
may easily present our results for a range of different atomic species,
and it is trivial to to apply a more accurate form of $V_{0}(\rho)$
given the appropriate data from theory~\citep{Nazin2008} or experiment~\citep{Evans2005}.

For Ps we also include an additional term~\citep{Stepanov2002} in
the local potential $V_{ps}(\bm{r})=V_{0}(\rho(\bm{r}))-V_{\epsilon}(\rho(\bm{r}))$
where $V_{\epsilon}=\frac{E_{ps}^{0}}{\epsilon_{r}^{2}}$, $E_{ps}^{0}=13.6\,\mathrm{eV}/4$
is the binding energy of Ps in vacuum and $\epsilon_{r}=\epsilon_{r}^{\infty}(\rho)$
is the high-frequency relative permittivity of the fluid.

\begin{figure}
\centering{}\includegraphics[width=0.24\textwidth]{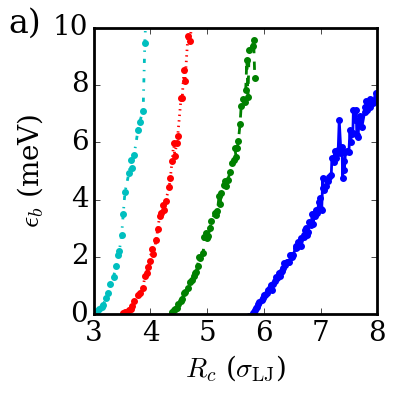}\includegraphics[width=0.24\textwidth]{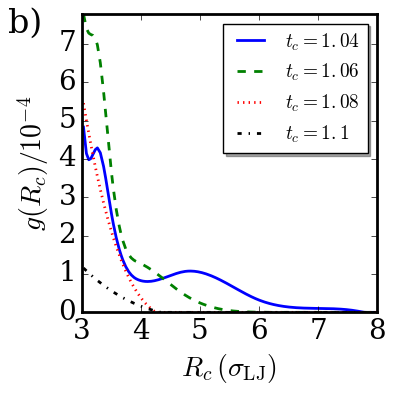}\caption{\label{fig:Binding-energies}The a) binding energies, $\epsilon_{b}$,
and b) density of states, $g(R_{c})$, for $e^{-}$ within high-density
fluctuations of argon at $T=1.0\,T_{\mathrm{LJ}}$ and $\rho=0.71\,\rho_{\mathrm{LJ}}$
for different identification thresholds $t_{c}$ and as a function
of fluctuation size $R_{c}$.}
\end{figure}

Performing an $s$-wave scattering calculation with the local potential
$V(r)$ results in scattering wavefunctions $\psi_{\mathrm{sc}}(r)$
and bound states $\psi_{b}(r)$ with binding energies $\epsilon_{b}$,
shown in figure~\ref{fig:Binding-energies}~a). We note that, at
these energies, higher partial waves do not contribute as the centrifugal
energy at the edge of the bubble $\hbar^{2}/2mR_{c}^{2}\geq20\,\mathrm{meV}$
is comparable to the thermal energy, and only low energy LPs are strongly
coupled to the bound states. To confirm this, we have calculated
the $p$-wave contribution for capture of $e^{-}$, $e^{+}$ and Ps
in argon and found it to be less than 1\% of the $s$-wave contribution.
 The same arguments also apply for bound states with $l\geq1$.

The coupling between the scattering and bound state which depends
spatially on the LP wavefunction and fluid density. We consider scattering
of the LP from sound mode phonons that provide the necessary transfer
of energy and momentum for capture into the fluctuations. As we show
in the supplementary material {[}{]}, attaining the appropriate transition
rate from quasi-free to bound states using Fermi's golden rule results
in a capture rate given by
\begin{align}
W(\epsilon\to\epsilon_{b})=\frac{S(0)}{4}\sqrt{\frac{2}{m}}\int dr\,\psi_{\epsilon}(r)j_{0}(\Delta k\,r)\psi_{\epsilon_{b}}(r)\nonumber \\
\times\rho(r)\sqrt{\tilde{\epsilon}}\sigma_{\mathrm{atom}}(\tilde{\epsilon}).\label{eq:transition-rate-general}
\end{align}
where $j_{l}(x)$ is the regular spherical Bessel function, $\tilde{\epsilon}=\max(\epsilon-\tilde{V}(r),0)$
and $\Delta k=c_{s}|\Delta\epsilon|$ where $c_{s}$ is the speed
of sound. Equation~(\ref{eq:transition-rate-general}) can be roughly
understood as an overlap integral between the scattering and bound
states, $\psi_{\epsilon}(r)$ and $\psi_{\epsilon_{b}}(r)$ respectively,
where momentum $\Delta k$ is transferred corresponding to the sound
mode dispersion relation and the coupling is dependent on the local
cross-section for a collision producing a sound wave, $\frac{S(0)}{4}\sqrt{\tilde{\epsilon}}\sigma_{\mathrm{atom}}(\tilde{\epsilon})$,
as well as the local atomic density, $\rho(r)$. Due to conservation
of momentum and energy, only low-energy LPs can be captured by the
flucutation.

The transition rate, $W(\epsilon\to\epsilon_{b})$ defines a cross-section
for capture $\sigma_{\mathrm{cap}}^{\{R_{c},t_{c}\}}(\epsilon)$ in
a collision with a single fluctuation. The overall capture rate, expressed
as a collision frequency, is then found by integrating over the density
of states for all available fluctuations:
\begin{equation}
\nu_{\mathrm{cap}}(\epsilon)=\sqrt{\frac{2\epsilon}{m}}\sum_{t_{c}}\int dR_{c}\,\sigma_{\mathrm{cap}}^{\{R_{c,}t_{c}\}}(\epsilon)g_{t_{c}}(R_{c}).\label{eq:cap_collfreq}
\end{equation}

\begin{figure}
\begin{centering}
\includegraphics[width=0.24\textwidth]{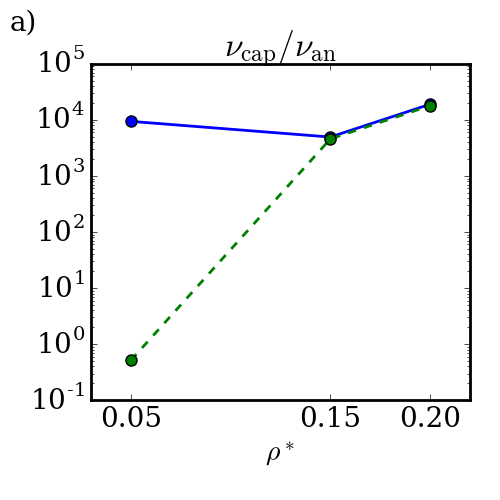}\includegraphics[width=0.24\textwidth]{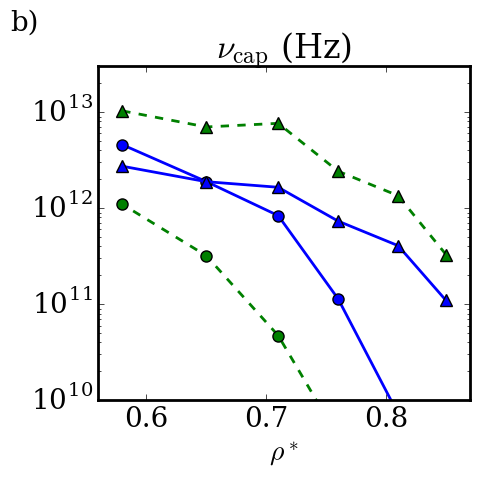}
\par\end{centering}

\caption{\label{fig:capture_rates}Thermally averaged capture rate $\nu_{\mathrm{cap}}$
for a) $e^{+}$ in the gas phase, and b) $e^{-}$ and Ps in the liquid
phase for argon (solid blue) and neon (dashed green). Ps curves are
distinguished by triangular markers. The $e^{+}$ capture rate is
compared to the annihilation rate $\nu_{\mathrm{an}}$.}
\end{figure}

\begin{table}
\begin{tabular}{cccccc}
 &  &  &  &  & \tabularnewline
\midrule
\midrule 
 &  & $\nu_{\mathrm{cap}}^{\mathrm{th}}$ ($\times10^{12}$~Hz) & Exp ($\rho\uparrow$) & Ref & Agreement\tabularnewline
\midrule
$e^{-}$ & Ne & EI zero~$\rightarrow1$ & $\mu\downarrow$, $\rho_{\mathrm{LJ}}\apprge0.20$ & \citep{Borghesani1990} & Possible\tabularnewline
 & Xe & AL $1\rightarrow10$ & $\mu\downarrow$, all $\rho_{\mathrm{LJ}}$ & \citep{Huang1978} & Good\tabularnewline
\midrule
\midrule 
 &  & $\nu_{\mathrm{cap}}/\nu_{\mathrm{an}}$ &  &  & \tabularnewline
\cmidrule{3-3} 
$e^{+}$ & Ar & AL $\approx10^{3}$ & $Z_{\mathrm{eff}}\uparrow$, all $\rho_{\mathrm{LJ}}$ & \citep{Tuomisaari1985} & Good\tabularnewline
 & Ne & EI zero~$\rightarrow10^{4}$ & $Z_{\mathrm{eff}}$ const & \citep{Canter1975a} & Good{*}\tabularnewline
 & Xe & AL $\approx10^{2}$ & $Z_{\mathrm{eff}}\uparrow$, $\rho_{\mathrm{LJ}}\rightarrow0$ & \citep{Tuomisaari1999} & Good\tabularnewline
\midrule
Ps & Ar & EI $10^{4}\rightarrow10^{5}$ & $Z_{\mathrm{eff}}^{1}\downarrow$ & \citep{Tuomisaari1985} & Good\tabularnewline
 & Ne & EI $10^{4}\rightarrow10^{6}$ & $Z_{\mathrm{eff}}^{1}\downarrow$, $\rho_{\mathrm{LJ}}\apprge0.10$ & \citep{Canter1975a} & Good{*}\tabularnewline
 & Kr & EI $10^{4}\rightarrow10^{5}$ & N/A &  & \tabularnewline
 & Xe & EI $10^{3}\rightarrow10^{4}$ & $Z_{\mathrm{eff}}^{1}\downarrow$, $\rho_{\mathrm{LJ}}\geq0.15$ & \citep{Tuomisaari1999} & Good{*}\tabularnewline
\midrule
\midrule 
 &  &  &  &  & \tabularnewline
\end{tabular}

\caption{\label{tab:gas_results}Comparison of thermally averaged $\nu_{\mathrm{cap}}^{\mathrm{th}}$
to experiments in dense gases. For $e^{-}$, a baseline of $\nu_{\mathrm{cap}}=10^{12}$~Hz
is used, see main text for details, and is compared to the experimental
mobility $\mu$. For $e^{+}$ and Ps, $\nu_{\mathrm{cap}}$ is compared
to direct annihilation $\nu_{\mathrm{an}}$. Shorthand notation of
EI (exponential increase) and AL (always large) have been used to
qualitatively describe the theoretical results. The trend of the experimental
observations is indicated for increasing density as the gas approaches
the gas-liquid phase transition or the critical density. An asterisk
indicates that time evolution of the fluctuation is required to explain
the good agreement.}
\end{table}

\begin{table}
\begin{tabular}{cccccc}
 &  &  &  &  & \tabularnewline
\midrule
\midrule 
 &  & $\nu_{\mathrm{cap}}^{\mathrm{th}}$ ($\times10^{12}$~Hz) & Exp ($\rho\downarrow$) & Ref & Agreement\tabularnewline
\midrule
$e^{-}$ & Ne & EI zero $\rightarrow1$ & SM near CP & \citep{Sakai1992} & Good\tabularnewline
 & Xe & AL $1\rightarrow10$ & $\mu\uparrow$ & \citep{Huang1978} & Possible\tabularnewline
\midrule
\midrule 
 &  & $\nu_{\mathrm{cap}}/\nu_{\mathrm{an}}$ &  &  & \tabularnewline
\cmidrule{3-3} 
$e^{+}$ & Ar & AL~$\approx10^{4}$ & $Z_{\mathrm{eff}}$ constant & \citep{Tuomisaari1985} & Bad\tabularnewline
 & Ne & Only $T^{*}=1.2$  & N/A &  & N/A\tabularnewline
 & Xe & Const~$\approx1$ & $Z_{\mathrm{eff}}$ constant & \citep{Tuomisaari1999} & Good\tabularnewline
\midrule
Ps & Ar & EI $10^{3}\rightarrow10^{4}$ & $Z_{\mathrm{eff}}^{1}\downarrow$ & \citep{Tuomisaari1985} & Good\tabularnewline
 & Ne & EI $10^{3}\rightarrow10^{4}$ & Bubbles in ACAR & \citep{Varlashkin1971} & Good\tabularnewline
 & Kr & EI $10^{3}\rightarrow10^{4}$ & Bubbles in ACAR & \citep{Varlashkin1971} & Good\tabularnewline
 & Xe & EI $10^{1}\rightarrow10^{3}$ & $Z_{\mathrm{eff}}^{1}\downarrow$  & \citep{Tuomisaari1999} & Good\tabularnewline
\midrule
\midrule 
 &  &  &  &  & \tabularnewline
\end{tabular}

\caption{\label{tab:liq_results}Summary of results with comparison to liquid
experiments. Details are as table~\ref{tab:gas_results}. Note, however,
that the experimental trend is given as density \emph{decreases }towards
the gas-liquid phase transition or critical density. The shorthand
``SM near CP'' stands for ``Scher-Montroll behavior near the critical
point''.}
\end{table}

\emph{Results in the noble gases and liquids---}We determine the fluctuation
properties $\rho_{\mathrm{fluc}}^{(R_{c},t_{c})}(r)$ and $g_{t_{c}}(R_{c})$
and perform the integrations (\ref{eq:transition-rate-general})~and~(\ref{eq:cap_collfreq})
using LJ parameters for neon, argon, krypton and xenon. Six points
on the liquid side of the gas-liquid coexistence region, spanning
$0.58\leq\rho_{\mathrm{LJ}}\leq0.85$ were chosen, as well as three
points in the gas phase, spanning $0.05\leq\rho_{\mathrm{LJ}}\leq0.20$
at constant temperature $T_{\mathrm{LJ}}=1.3$. Note that the critical
temperature and density for the untruncated LJ fluid are $\rho_{\mathrm{LJ},c}\approx0.3$
and $T_{\mathrm{LJ},c}\approx1.33$. We used the static structure
factor $S(0)$ calculated from the pair correlator of our Monte-Carlo
simulations for each temperature and density. 

As shown in tables~\ref{tab:gas_results} and~\ref{tab:liq_results},
our results agree favorably with most of the experimental data we
have found. This is remarkable, given the gas scattering-length approximation
we have made for $V_{0}$. In general, we try to compare the thermally
averaged capture rate, $\nu_{\mathrm{cap}}^{\mathrm{th}}$, to relevant
time scales of the transport. For $e^{+}$ and Ps this is naturally
represented by its ratio to the annihilation frequency $\nu_{\mathrm{an}}$,
as capture must be fast enough to observe an increase in total annihilation
rate, more commonly described as an increase in the effective electron
number per atom, $Z_{\mathrm{eff}}$ and $Z_{\mathrm{eff}}^{1}$ for
$e^{+}$ and Ps respectively.

For $e^{-}$ a natural time scale for comparison is the transit time,
which is unfortunately different for each experimental configuration.
However, we have found that assuming noticeable capture frequencies
must be $10^{12}$~Hz or larger gives very good agreement with experimental
observations (an improved comparison requires a full transport calculation).
In neon, we predict that as the density in the gas (liquid) phase
is increased (decreased), capture becomes exponentially more likely.
This follows the observation of ``activation'' densities in experiment,
where modified mobilities $\mu$~\citep{Borghesani1990} or Scher-Montroll
behavior~\citep{Sakai1992} occur for a limited range of densities
about the critical point. For xenon, we predict a strong capture rate
for all of the densities we investigated, which agrees well for experiments
in the gas phase~\citep{Huang1978}. In the liquid phase there is
possible agreement, but the experimental observations are overwhelmed
by an increase in mobility which is likely due to a non-linear change
of $V_{0}$ to repulsive behavior \citep{Evans2005}, neglected in
our scattering length approximation.

For $e^{+}$ in gases, our predictions of strong capture rates agree
well with experiments in argon~\citep{Tuomisaari1985} and xenon~\citep{Tuomisaari1999}.
In contrast, experiments in neon~\citep{Canter1975a} see no increased
annihilation which appear to contradict our results. However, this
can be explained through preliminary calculations of the Navier-Stokes
evolution of the high-density fluctuation into a cluster, see supplementary
material {[}{]}. Because the $e^{+}$-Ne interaction is weak in comparison
to the other noble gases, the bound state is quickly lost in time
evolution before a significant change in fluid density can occur.
In the liquid phase, we find agreement with xenon measurements, whereby
a capture rate is obtained that is too slow to modify the total annihilation
rate. Unfortunately, our prediction of a strong capture rate in argon
liquid is at odds with experimental observations \citep{Tuomisaari1985}
and is not easily explained through our preliminary time evolution
calculations. We conjecture that non-linearities in $V_{0}$ play
a dominant role in this case.

In general, Ps has been observed to form bubbles in all noble-gas
liquids~\citep{Varlashkin1971,Tuomisaari1985,Tuomisaari1999} and
for high-density gases~\citep{Canter1975a}, in agreement with our
calculations. However, our results also predict that strong capture
rates persist even to relatively dilute gases. The apparent discrepancy
can easily be explained by identifying the wide and shallow profiles
corresponding to these fluctuations, which do not allow the formation
of stable bubbles in the subsequent time-evolution.

\emph{Conclusion}---We have developed a theory, valid in both dense
gases and liquids, which describes scattering of light particles into
large-scale natural fluctuations and the formation of trapped states.
The theory has no fitting parameters and agrees well with almost all
of the experimental literature in the noble gases and liquids. This
work represents the first calculation of $\sigma_{\mathrm{cap}}(\epsilon)$
which will be used in upcoming transport calculations that are far
from equilibrium, combined with a model of the time-dependence of
the fluctuations that describes the waiting time distribution for
the localized states~\citep{Philippa2014,Stokes2015}. The strongest
approximation in the calculations presented here is the use of a gas
scattering-length which can be improved by considering Wigner-Seitz
models \citep{Evans2005}. Further investigations will also be extended
to include non-local contributions (e.g. including spatial dependence
of screening) which would also provide a natural pathway to describing
polar molecules.

\bibliography{flucap1}

\appendix
\begin{table*}
\begin{tabular}{cccccccc}
\hline 
Atom & $\epsilon_{LJ}$~($k_{B}/K$) & $r_{LJ}\,(\angstrom)$ & $c_{s}$~m/s & $e^{-}$ $a_{s}$ $(a_{0})$ & $e^{+}$ $a_{s}$ $(a_{0})$ & Ps $a_{s}$ $(a_{0})$ & $K_{CM}$ (cm$^{3}$/mole)\tabularnewline
\hline 
\hline 
Ne & $37.29$~\citep{Ramirez2008} & $2.782$ & 540~\citep{KayeLaby} & $0.215$~\citep{LandBorn97} & $-0.467$~\citep{Green2014} & $1.56$~\citep{Mitroy2003} & $2.290$ \tabularnewline
\hline 
Ar & $119.8$~\citep{Wang2005} & $3.405$ & 813~\citep{Lide2004} & $-1.452$~\citep{LandBorn97} & $-4.41$~\citep{Green2014} & $2.14$~\citep{Fabrikant2014} & $4.106$~\citep{Amey1964}\tabularnewline
\hline 
Kr & $164.4$~\citep{Wang2005} & $3.638$ & 1120~\citep{Rabinovich1988} & $-3.35$~\citep{LandBorn97} & $-9.71$~\citep{Green2014} & $2.35$~\citep{Fabrikant2014} & $6.239$~\citep{Amey1964}\tabularnewline
\hline 
Xe & $231.1$~\citep{Wang2005} & $3.961$ & 1090~\citep{Rabinovich1988} & $-6.3$~\citep{LandBorn97} & $-84.5$~\citep{Green2014} & $2.29$~\citep{Mitroy2003} & $9.685$~\citep{Amey1964}\tabularnewline
\hline 
\end{tabular}

\caption{\label{tab:liq_params}Parameters used for each of the noble gas species
in the calculations. $K_{CM}$ is the Claussius-Mossotti coefficient,
defined by $\epsilon_{r}(\rho)=\frac{1+2K_{CM}\rho}{1-K_{CM}\rho}$.
The LJ parameters for argon, krypton and xenon have been taken from
the ``Horton'' values listed in \citep{Wang2005}. <Citations for
all numbers!>}
\end{table*}

\section{Details of Fluctuation Identification}

We include some additional details of the identification procedure
here. As described in the main text, the identification of fluctuations
occurs by 1) Gaussian blurring, 2) classification of fluctuations
identified by an average in a spherical volume of radius $R_{c}$
less than a threshold $t_{c}$, 3) obtaining the density distribution
for each fluctuation.

The Gaussian blurring is done through representing a continuum density
distribution on a fine grid, replacing each discrete atom by a Gaussian
of width $\sigma_{\mathrm{blur}}=(3/4\pi\rho_{0})^{1/3}$:
\begin{equation}
\rho(\bm{x})=\sum_{i=0}^{N}\frac{1}{(2\pi\sigma_{\mathrm{blur}}^{2})^{3/2}}e^{-|\bm{x}_{i}-\bm{x}|_{L}^{2}/2\sigma_{\mathrm{blur}}^{2}}
\end{equation}
where $|\ldots|_{L}$ indicates the smallest separation between the
two points accounting for periodic boundaries. We make sure that each
atom is individually normalized on the grid before adding it to the
total density distribution.

Next, the density distribution is averaged in a sphere of radius $R_{c}$
around each grid point. The radii are taken to be successively larger
and, when searching for bubbles, we choose to identify these regions
as connected groups of grid points whose average in a radius $R_{c}$
is smaller than a given threshold $t_{c}$ relative to the homogeneous
density, $\rho_{0}$, i.e. $\rho_{\mathrm{avg}}<t_{c}\rho_{0}$. These
groups are identified using the Hoshen-Kopelman cluster-identification
algorithm~\citep{Hoshen1976}. The process is identical for high-density
fluctuations except that the condition is for the average to be larger
than the given threshold $t_{c}$, i.e. $\rho_{\mathrm{avg}}>t_{c}\rho_{0}$.
For thresholds close to unity or for small radii, there may be many
connected groups that span a large number of grid points. For example,
a weak threshold $t_{c}=1.0001$ would likely identify a group of
points which percolate across the entire system. However, as the threshold
or the radius is increased, each connected group becomes smaller and
more defined until the point at which it disappears. As we only wish
to identify the location of the group itself, we look for these points
where a group disappears on increase of $R_{c}$ or $t_{c}$. Practically
this is done by filtering out overlapping fluctuations from a sequence
of identifications done for different $R_{c}$ values at a fixed threshold.

We are only interested in fluctuations that are larger than the average
spacing between molecules, so we begin the $R_{c}$ scan at $R_{c,\mathrm{min}}=3$.
The identification process should be reasonably insensitive to the
choice of $R_{c,\mathrm{min}}$, so long as it remains small enough
to allow for the larger fluctuations of interest. Changes in $R_{c,\mathrm{min}}$
should be reflected as changes in the $t_{c}$ assigned to each fluctuation.

The location for each fluctuation identified in this manner can be
taken to be the center of the group of grid points, found through
the mean position of all grid points in the group. For periodic systems,
such a mean can be ambiguous (e.g. when the group spans a boundary
of the system) however our case is much simpler, as each group is
comprised of only a few connected points. Practically, we identify
the fluctuation position as the grid point in the group which minimizes
the sum of distances to all other grid points in the group.

\section{Capture Coupling Strength}

The capture cross section described in the main text results from
a coupling between the incoming scattering state and a bound state
in the fluctuation. We use Fermi's golden rule to obtain an approximate
coupling and fix the prefactor by a free-to-free scattering event.
We assume the coupling itself is a result of the collision of the
incoming light-particle with a sound mode, which is necessary to produce
the energy transfer required. Hence, we postulate that the time-dependent
coupling in a homogeneous fluid is of the form 
\begin{equation}
V_{\mathrm{coup}}(\bm{x},t)=C(\Delta\epsilon)e^{i(\Delta\bm{k}\cdot\bm{x}-\Delta\epsilon t/\hbar)}.\label{eq:coupling_form}
\end{equation}
The function $C(\Delta\epsilon)$ can be found by comparing to a free-to-free
scattering event with that of the Boltzmann equation, resulting in:
\begin{equation}
C_{\mathrm{homo}}(\Delta\epsilon)=\rho_{0}\,\sqrt{\frac{2}{m}}\int d\Delta k\,\sigma_{\mathrm{sound}}(\Delta k,\Delta\epsilon).
\end{equation}
The cross section for a collision with a sound mode, $\sigma_{\mathrm{sound}}(\Delta k,\Delta\epsilon)$,
is approximated for small energy transfer by its form in the hydrodynamic
limit \citep{Hansen1976} and neglecting sound wave attenuation, $\sigma_{\mathrm{sound}}(\Delta k,\Delta\epsilon)\approx\frac{1}{4}S(0)\delta(\Delta\epsilon+c_{s}\Delta k)\sigma_{\mathrm{atom}}(\epsilon)$.
Here, $\sigma_{\mathrm{atom}}(\epsilon)$ is the atomic cross-section
which we approximate at these low energies by$\sigma_{\mathrm{atom}}=4\pi a_{s}^{2}$
where $a_{s}$ is the scattering length, with the explicit values
used listed in table~\ref{tab:liq_params}. The factor of $1/4$
arises from our interest in only the creation of sound modes and not
collisions from existing sound waves, as well as assuming the adiabatic
ratio is $\gamma\approx2$. This leaves us with: 
\begin{equation}
C_{\mathrm{homo}}(\Delta\epsilon)=\frac{\hbar}{2\pi}\frac{S(0)}{4}\sqrt{\frac{2}{m}}\rho_{0}\sqrt{\epsilon}\sigma_{\mathrm{atom}}(\epsilon).
\end{equation}
In the non-homogeneous case of a fluctuation, we make a simple local
approximation to the velocity, substituting $\epsilon\rightarrow\tilde{\epsilon}=\max(\epsilon-\tilde{V}(r),0)$,
resulting in:
\begin{equation}
C(\Delta\epsilon)=\frac{\hbar}{2\pi}\frac{S(0)}{4}\sqrt{\frac{2}{m}}\rho(\bm{r})\sqrt{\tilde{\epsilon}}\sigma_{\mathrm{atom}}(\tilde{\epsilon}).\label{eq:coeff_inhomo}
\end{equation}
Finally, substituting (\ref{eq:coeff_inhomo}) and (\ref{eq:coupling_form})
into Fermi's golden rule, and then integrating over angles results
in equation (\ref{eq:transition-rate-general}).

\section{Preliminary Time Evolution Results}

To investigate the time-dependent behavior of the fluctuation after
capture of the light particle, we perform a crude hydrodynamic evolution
of the fluid density using the Navier-Stokes equations. Although the
stable bubble or cluster is a microscopic system, where the validity
of hydrodynamic limit is highly questionable, we are mostly interested
in the initial behavior of the fluctuation which is at a mesoscopic
scale.

The Navier-Stokes equations we use assume spherical symmetry and hence
take the form:
\begin{gather}
\frac{\partial\rho}{\partial t}=-\frac{1}{r^{2}}\frac{\partial}{\partial r}\left(r^{2}\rho(r,t)u(r,t)\right)\label{eq:dens_continuuity}\\
\begin{aligned}\frac{\partial u}{\partial t}+u\frac{\partial u}{\partial r}= & -\frac{\nabla P(\rho)}{\rho}+\frac{\nu(\rho)}{\rho}\frac{\partial}{\partial r}\left(r^{2}\frac{\partial u}{\partial r}\right)\\
 & +\frac{1}{3}\nu\nabla(\frac{1}{r^{2}}\frac{\partial(r^{2}u)}{\partial r^{2}})-\frac{\nabla V(\rho,\psi)}{\rho}
\end{aligned}
\label{eq:flow_continuuity}
\end{gather}
where $u(r,t)$ is the flow velocity, $P(\rho)$ is the local pressure
obtained from our Monte-Carlo simulations at varying densities and
$\nu(\rho)$ is the viscosity given for the LJ fluid in \citep{Meier2012}.
The force on the fluid results from the potential $V$, which depends
on the fluid/light-particle interaction given by $V(\rho,\psi)=V_{0}(\rho)|\psi|^{2}$
where $V_{0}$ is defined in the main text.

We assume the wavefunction adiabatically follows the fluid density
profile and so integrate equations~(\ref{eq:dens_continuuity}) and~(\ref{eq:flow_continuuity})
in time, regularly recalculating the wavefunction after small time
intervals. We continue the integration until either the binding energy
of the light particle stabilizes, indicating a stable dressed particle
has been reached, or until the profile no longer supports a bound
state, in which case the light particle has been ``popped'' from
the fluctuation and is assumed to return to a quasi-free state.

We note that a simple estimate of the stability of the bubble in the
dilute gas limit can be obtained by comparing the forces due to pressure
and the light-particle interaction. As $P\propto\rho$ and $V_{0}(\rho)\propto\rho$
for dilute gases and the wavefunction density decreases with size
of fluctuation $|\psi|^{2}\propto1/R_{c}$, then the ratio $V/P\propto1/R_{c}$.
This means that the fluid/light-particle interaction becomes less
important the larger the fluctuations and hence the more dilute the
gas.

\end{document}